\documentclass[a4paper,12pt]{article}  

\usepackage{graphicx}
\usepackage{amssymb}

\title{A new approach to the Thomas-Fermi boundary-value problem}

\date{\today}

\author{Giampiero Esposito ORCID: 0000-0001-5930-8366 \\
Istituto Nazionale di Fisica Nucleare, Sezione di
Napoli, \\ 
Complesso Universitario di Monte S. Angelo, \\ 
Via Cintia Edificio 6, 80126 Napoli, Italy \and
Salvatore Esposito  ORCID: 0000-0003-3099-5574 \\
Istituto Nazionale di Fisica Nucleare, Sezione di
Napoli, \\ 
Complesso Universitario di Monte S. Angelo, \\ 
Via Cintia Edificio 6, 80126 Napoli, Italy}

\begin{document}
\maketitle

\begin{abstract}
Given the Thomas-Fermi equation
$\sqrt{x}\varphi''=\varphi^{3 \over 2}$, this paper 
changes first the dependent variable by defining
$y(x) \equiv \sqrt{x \varphi(x)}$. The boundary
conditions require that $y(x)$ must
vanish at the origin as $\sqrt{x}$, 
whereas it has a fall-off behaviour 
at infinity proportional to the power 
${1 \over 2}(1-\chi)$ of the independent variable $x$,
$\chi$ being a positive number. Such boundary
conditions lead to a $1$-parameter family of approximate solutions
in the form $\sqrt{x}$ times a ratio of finite linear combinations
of integer and half-odd powers of $x$. If $\chi$ is set
equal to $3$, in order to agree exactly with the asymptotic solution
of Sommerfeld, explicit forms of the 
approximate solution are obtained for all values of $x$.
They agree exactly with the Majorana solution at small $x$,
and remain very close to the numerical solution for 
all values of $x$. Remarkably, without making any use of
series, our approximate solutions achieve a smooth transition
from small-$x$ to large-$x$ behaviour. Eventually, 
the generalized Thomas-Fermi equation that includes
relativistic, non-extensive and thermal effects is studied,
finding approximate solutions at small and large $x$ for small or
finite values of the physical parameters in this equation.
\end{abstract}

\section{Introduction}
\setcounter{equation}{0}

Since the early days of quantum mechanics, it was of interest
to investigate a hybrid model where the electrostatic potential
$V$ due to the nucleus and to the cloud of electrons obeys
again a Poisson equation but with a charge density that is
affected by quantum mechanics \cite{Thomas,Fermi}. 
On assuming a central potential, one can write
\begin{equation}
V(r)=\varphi(r){Ze \over r},
\label{(1.1)}
\end{equation}
where $\varphi(r)$ is the ratio between the effective atomic
number $Z_{\rm eff}$ and the atomic number $Z$, and is the
function describing how the mutual repulsion of electrons 
modifies the otherwise Coulomb-type potential 
${Ze \over r}$. The potential $V(r)$ is required to approach
the pure Coulomb form as $r \rightarrow 0$, while it has to
vanish as $r \rightarrow \infty$, in order to ensure that
the atom as a whole is uncharged. Eventually, one arrives at
the Thomas-Fermi boundary-value problem, consisting of a
non-linear equation that, in dimensionless units, 
reads as \cite{Majorana,EspositoS}
\begin{equation}
\sqrt{x}{d^{2}\varphi \over dx^{2}}=\varphi^{3 \over 2},
\label{(1.2)}
\end{equation}
supplemented by the boundary conditions at the origin and
at infinity (cf. Ref. \cite{EspositoS})
\begin{equation}
\lim_{x \to 0}\varphi(x)=1,
\label{(1.3)}
\end{equation}
\begin{equation}
\lim_{x \to \infty}\varphi(x)=0.
\label{(1.4)}
\end{equation}
The aim of the present paper is to develop a new method for
solving the Thomas-Fermi boundary-value problem, that relies
on a more convenient form of Eq. (1.2) and a more careful 
formulation of the boundary condition at infinity. For this 
purpose, section $2$ studies a change of dependent variable 
and the resulting equations. Section $3$ considers a
$1$-parameter family of boundary conditions at infinity,
while section $4$ obtains approximate solutions of the problem 
(1.2)-(1.4) by means of a ratio of linear combinations of integer
and half-odd powers of $x$. Plots of approximate vs. numerical 
solutions are displayed in section $5$. Section $6$ is instead 
devoted to solving the generalized Thomas-Fermi equation that
includes relativistic, non-extensive and thermal corrections.
Concluding remarks are presented in section $7$.

\section{Ch\-an\-ge of de\-pen\-de\-nt va\-ria\-ble fo\-r the 
Tho\-mas-Fer\-mi eq\-ua\-ti\-on}
\setcounter{equation}{0}

In Eq. (1.2), fractional powers of $x$ and $\varphi$ are an
undesirable feature if one wants to deal with integer powers of
the unknown function and its derivatives, but if we multiply
both sides by $\sqrt{\varphi}$ we obtain
\begin{equation}
\sqrt{x \varphi}{d^{2}\varphi \over dx^{2}}=\varphi^{2}
={\varphi^{2}x^{2}\over x^{2}}.
\label{(2.1)}
\end{equation}
This suggests defining
\begin{equation}
y \equiv \sqrt{x \varphi(x)},
\label{(2.2)}
\end{equation}
which, by virtue of (1.3), implies the boundary condition at
the origin
\begin{equation}
\lim_{x \to 0} \Bigr[x^{-{1 \over 2}}y(x)\Bigr]=\pm 1.
\label{(2.3)}
\end{equation}
Moreover, by virtue of the definition (2.2), we obtain
$\varphi(x)={y^{2}(x)\over x}$, and hence Eq. (2.1) reads as
\begin{equation}
y {d^{2}\over dx^{2}} \left({y^{2}\over x}\right)
={y^{4}\over x^{2}},
\label{(2.4)}
\end{equation}
i.e.
\begin{equation}
2{y^{3}\over x^{3}}-4{y^{2}\over x^{2}}{dy \over dx}
+2{y^{2} \over x}{d^{2}y \over dx^{2}}
+2{y \over x}\left({dy \over dx}\right)^{2}
={y^{4}\over x^{2}}.
\label{(2.5)}
\end{equation}
This suggests multiplying both sides of Eq. (2.5) by
${x \over 2y^{2}}$, obtaining therefore the quasi-linear
equation (i.e. linear with respect to the highest order
derivative)
\begin{equation}
\left[{d^{2}\over dx^{2}}-{2 \over x}{d \over dx}
+{1 \over x^{2}}\right]y={1 \over 2}{y^{2}\over x}
-{1 \over y} \left({dy \over dx}\right)^{2}.
\label{(2.6)}
\end{equation}
The operator on the left-hand side is a linear second-order
operator for which the origin is a regular singular
point. The non-linear terms occur on the right-hand side
of Eq. (2.6). 

\section{The boundary condition at infinity}
\setcounter{equation}{0}

The boundary condition at infinity needs a more careful 
formulation, since the rate of fall-off is not specified
by Eq. (1.4). The large-$x$ solution found by 
Sommerfeld \cite{Sommerfeld}, i.e.
$\varphi(x) \sim {144 \over x^{3}}$, fails to satisfy 
Eq. (1.3) and hence it is not a solution of the 
Thomas-Fermi boundary-value problem (1.2)-(1.4). However,
it remains of some value because it suggests considering
a positive number $\chi$ for which
\begin{equation}
\lim_{x \to \infty}x^{\chi}\varphi(x)={\rm constant}.
\label{(3.1)}
\end{equation}
By virtue of our definition (2.2), Eq. (3.1) can be
re-expressed in the form
\begin{equation}
\lim_{x \to \infty}x^{\chi -1}y^{2}(x)
={\rm constant},
\label{(3.2)}
\end{equation}
i.e.
\begin{equation}
\lim_{x \to \infty}\Bigr[x^{{1 \over 2}(\chi-1)}y(x)\Bigr]
=\sigma={\rm constant}.
\label{(3.3)}
\end{equation}

We also notice, by inspection of Eqs. (2.3) and (3.3), that
the boundary conditions of the Thomas-Fermi boundary-value
problem can be expressed in a unified way by a single formula:
\begin{equation}
\lim_{x \to a}\Bigr[x^{{1 \over 2}(f(a)-1)}y(x)\Bigr]=g(a),
\label{(3.4)}
\end{equation}
where
\begin{equation}
f(a=0)=0 \; \; \; \; \; \; g(a=0)=\pm 1,
\label{(3.5)}
\end{equation}
\begin{equation}
f(a=\infty)=\chi \; \; \; \; \; \; 
g(a=\infty)={\rm constant}.
\label{(3.6)}
\end{equation}
This is a simple but non-trivial feature, never noted before
to the best of our knowledge.

\section{A family of approximate solutions for all values of $x$}
\setcounter{equation}{0}

\subsection{Large-$x$ behaviour}

Since $\varphi(x)={144 \over x^{3}}={y^{2}(x)\over x}$ solves 
exactly Eq. (1.2) at large $x$, we know that the desired function
$y$ should approach ${12 \over x}$ at large $x$
(this is fixed up to a sign, but such a detail is inessential
for physical purposes). Moreover,
we know from the analysis of the boundary-value problem that
$y(x)$ should be dominated by $\sqrt{x}$ as 
$x \rightarrow 0$. Our task is therefore to look for a smooth
interpolation between such limiting behaviours. For this purpose,
we point out that power series are incompatible with both limiting
behaviours, whereas rational functions are incompatible only with
the $\sqrt{x}$ behaviour as $x \rightarrow 0$. These features
suggest considering ratios of linear combinations of integer and
half-odd powers of $x$, that we divide into four sets as follows.
\vskip 0.3cm
\noindent
{\bf Case $1$}. On denoting 
hereafter by $l$ and $m$ two positive integers, we can write
\begin{equation}
y_{1}(x)= \sqrt{x}
{\left[1+\alpha_{1}x^{1 \over 2}+a_{1}x+...
+\alpha_{l}x^{l-{1 \over 2}}+a_{l}x^{l}\right] \over
\left[1+\beta_{1}x^{1 \over 2}+b_{1}x+...
+\beta_{m}x^{m- {1 \over 2}}+b_{m}x^{m}\right]}.
\label{(4.1)}
\end{equation}
As $x \rightarrow \infty$, $y_{1}(x)$ approaches
$ {a_{l}\over b_{m}}
{1 \over x^{m-l-{1 \over 2}}}$. This case is therefore ruled out
because integer values of $l$ and $m$ are incompatible with
the condition
$$
m-l-{1 \over 2}=1
$$
that is enforced by the Sommerfeld solution at large $x$.
\vskip 0.3cm
\noindent
{\bf Case $2$}. Here $y(x)$ is taken to be
\begin{equation}
y_{2}(x)= \sqrt{x}
{\left[1+\alpha_{1}x^{1 \over 2}+a_{1}x+...
+\alpha_{l}x^{l-{1 \over 2}}\right] \over
\left[1+\beta_{1}x^{1 \over 2}+b_{1}x+...
+\beta_{m}x^{m- {1 \over 2}}\right]}.
\label{(4.2)}
\end{equation}
As $x \rightarrow \infty$, $y_{2}(x)$ approaches
${\alpha_{l}\over \beta_{m}}
{1 \over x^{m-l-{1 \over 2}}}$, which is therefore ruled
out for the same reason as in Case $1$.
\vskip 0.3cm
\noindent
{\bf Case $3$}. We consider $y(x)$ given by
\begin{equation}
y_{3}(x)= \sqrt{x}
{\left[1+\alpha_{1}x^{1 \over 2}+a_{1}x+...
+\alpha_{l}x^{l-{1 \over 2}}+a_{l}x^{l}\right] \over
\left[1+\beta_{1}x^{1 \over 2}+b_{1}x+...
+\beta_{m}x^{m- {1 \over 2}}\right]}.
\label{(4.3)}
\end{equation}
As $x \rightarrow \infty$, $y_{3}(x)$ approaches
${a_{l}\over \beta_{m}}
{1 \over x^{m-l-1}}$. This can equal 
${12 \over x}$ provided that 
\begin{equation}
{a_{l}\over \beta_{m}}=12, \; \; m-l-1=1.
\label{(4.4)}
\end{equation}
\vskip 0.3cm
\noindent
{\bf Case $4$}. Last, we can assume that
\begin{equation}
y_{4}(x)= \sqrt{x}
{\left[1+\alpha_{1}x^{1 \over 2}+a_{1}x+...
+\alpha_{l}x^{l-{1 \over 2}}\right] \over
\left[1+\beta_{1}x^{1 \over 2}+b_{1}x+...
+\beta_{m}x^{m- {1 \over 2}}+b_{m}x^{m}\right]}.
\label{(4.5)}
\end{equation}
As $x \rightarrow \infty$, $y_{4}(x)$ approaches
${\alpha_{l} \over b_{m}}
{1 \over x^{m-l}}$. This can equal ${12 \over x}$
provided that
\begin{equation}
{\alpha_{l}\over b_{m}}=12, \; \; m-l=1.
\label{(4.6)}
\end{equation}
\vskip 0.1cm
\noindent
Thus, only cases $3$ and $4$ are picked out by the requirement
of recovering the Sommerfeld behaviour at large $x$.

\subsection{Small-$x$ behaviour}

As $x \rightarrow 0$, Majorana \cite{Majorana,Digrezia} obtained a
formula in excellent agreement with the numerical solution.
According to his analysis, the solution of Eq. (1.2)
has the small-$x$ behaviour 
\begin{equation}
\varphi(x) \sim 1-px+{4 \over 3}x^{3 \over 2}
-{2 \over 5}p x^{5 \over 2}+{\rm O}(x^{2}),
\label{(4.7)}
\end{equation}
where $p \approx 1.58$ and the fourth term on the right-hand side
improves the previous analysis of Fermi \cite{Fermi}, who did
not go beyond $x^{3 \over 2}$. Hence we obtain, 
as $x \rightarrow 0$, 
\begin{equation}
y(x)=\sqrt{x \varphi(x)} \sim \sqrt{x}
\left[1-{1 \over 2}px+{2 \over 3}x^{3 \over 2}
-{1 \over 8}p^{2}x^{2}+{2 \over 15}p x^{5 \over 2}
+{\rm O}(x^{3})\right],
\label{(4.8)}
\end{equation}
where we have exploited the Taylor expansion of
$\sqrt{1+z}$ about $z=0$, having set
$$
z \equiv -px+{4 \over 3}x^{3 \over 2}
-{2 \over 5}p x^{5 \over 2}.
$$
Now we require that the approximate solutions (4.3) and (4.5),
when expanded about $x=0$, agree with Eq. (4.8). This can be
achieved with a patient calculation, leaving the numerator of
(4.3) and (4.5) untouched, while the inverse of the denominator
is expanded according to the geometric series algorithm for
$|w|<<1$, i.e.,
$$
{1 \over (1+w)} \sim 1-w+w^{2}-w^{3}+{\rm O}(w^{4}).
$$

As is clear from Eqs. (4.4) and (4.6), we can regard $l$ as
being freely specifiable, while
\begin{equation}
m=l+2 \; {\rm in} \; y_{3}, \; m=l+1
\; {\rm in} \; y_{4}.
\label{(4.9)}
\end{equation}
The allowed approximate solutions can be therefore denoted
by $y_{l,m}(x)$, where
\begin{equation}
y_{l,l+2}(x)=y_{3}(x), \;
y_{l,l+1}(x)=y_{4}(x).
\label{(4.10)}
\end{equation}
The labels $l$ and $m$ tell us explicitly that we have solved a
boundary-value problem, since their values affect our choice of
how many integer and half-odd powers of $x$ should occur in
(4.3) and (4.5) in order to fulfill the boundary conditions. 
For example, when $l$ is set to $1$ for simplicity,
the requirement that $y_{1,3}(x)$ should agree as 
$x \rightarrow 0$ with Eq. (4.8), leads to the following
values of the coefficients:
\begin{equation}
\alpha_{1}=\beta_{1}, \; a_{1}={p \over 30}
(-32+15 p \beta_{1}),
\label{(4.11)}
\end{equation}
\begin{equation}
b_{1}={p \over 30}(-17+15p \beta_{1}), \;
\beta_{2}={1 \over 6}(-4+3p \beta_{1}),
\label{(4.12)}
\end{equation}
\begin{equation}
b_{2}={1 \over 120}(-19p^{2}-80 \beta_{1}+30 p^{3}\beta_{1}),
\label{(4.13)}
\end{equation}
\begin{equation}
\beta_{3}={p \over 360}(-32+15p \beta_{1}),
\label{(4.14)}
\end{equation}
where we note that
\begin{equation}
a_{1}=12 \beta_{3}
\label{(4.15)}
\end{equation}
in order to agree with the Sommerfeld condition
$y_{1,3}(x) \sim  {12 \over x}$ as $x \rightarrow \infty$.
In particular, upon setting $\beta_{1}=0$, we find
\begin{equation}
y_{1,3}(x)=  \sqrt{x} {\left(1-{16 \over 15}px \right) \over
\left(1-{17 \over 30}px-{2 \over 3}x^{3 \over 2}
-{19 \over 120}p^{2}x^{2}-{4 \over 45}p x^{5 \over 2}\right)},
\label{(4.16)}
\end{equation}
and, with entirely analogous procedure,
\begin{equation}
y_{1,2}(x)= \sqrt{x} {\left[1-{p \over 2}x
+\left({p^{3}\over 6}+{8 \over 9}\right)x^{3 \over 2}\right]
\over
\left[1+\left({p^{3}\over 6}+{2 \over 9}\right)x^{3 \over 2}
+{p^{2}\over 8}x^{2}+\left({p^{4}\over 12}-{p \over 45}\right)
x^{5 \over 2}+\left({p^{3}\over 72}+{2 \over 27}\right)x^{3}
\right]}.
\label{(4.17)}
\end{equation}

\section{Plots of $y(x)$ and $\varphi(x)$}
\setcounter{equation}{0}

Hereafter, we plot our approximate solutions $y_{1,3}(x)$ and
$y_{1,2}(x)$ with $\alpha_{1}=\beta_{1}=0$, against the numerical
solution of Eq. (2.6). Moreover, we also plot the resulting
approximate solutions of the Thomas-Fermi boundary-value problem
(1.2)-(1.4), i.e.,
\begin{equation}
\varphi_{1,3}(x)={[y_{1,3}(x)]^{2}\over x}, \;
\varphi_{1,2}(x)={[y_{1,2}(x)]^{2}\over x}.
\label{(5.1)}
\end{equation}
against the corresponding numerical solution. 

\begin{figure}
\centering
\includegraphics[height=8cm]{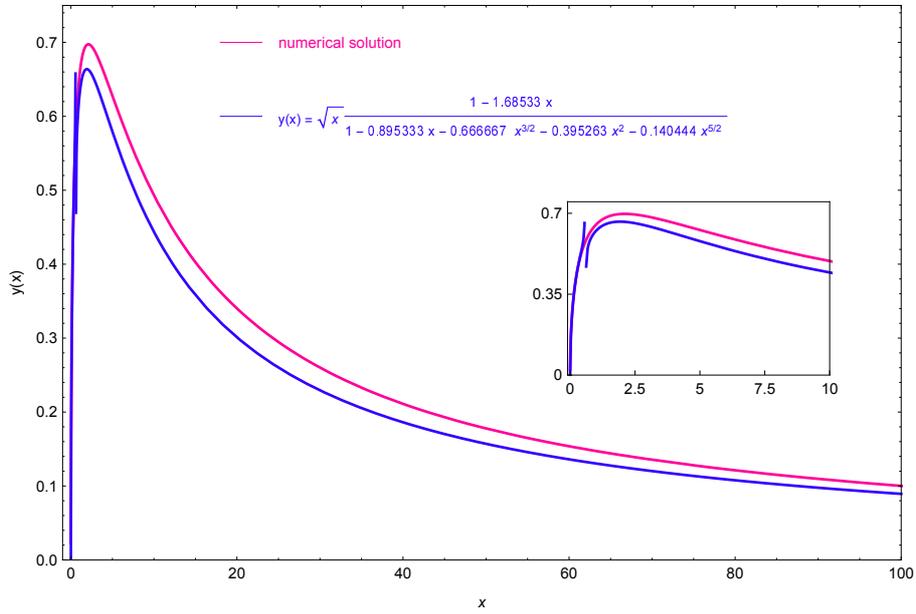}
\caption{Our $y_{1,3}(x)$ in Eq. (4.16) vs. the numerical solution.}
\end{figure}

\begin{figure}
\centering
\includegraphics[height=8cm]{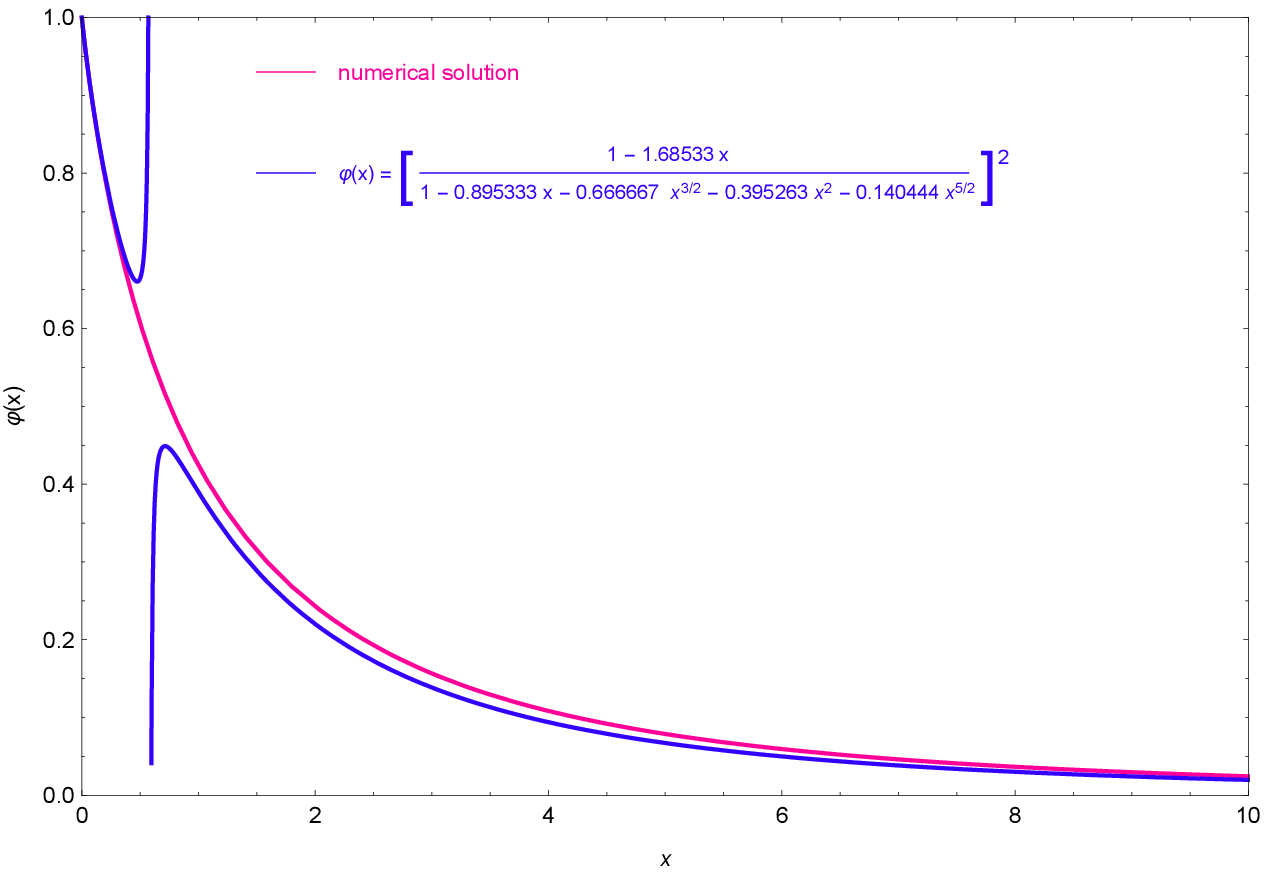}
\caption{Our $\varphi_{1,3}(x)$ vs. the numerical solution.}
\end{figure}

\begin{figure}
\centering
\includegraphics[height=8cm]{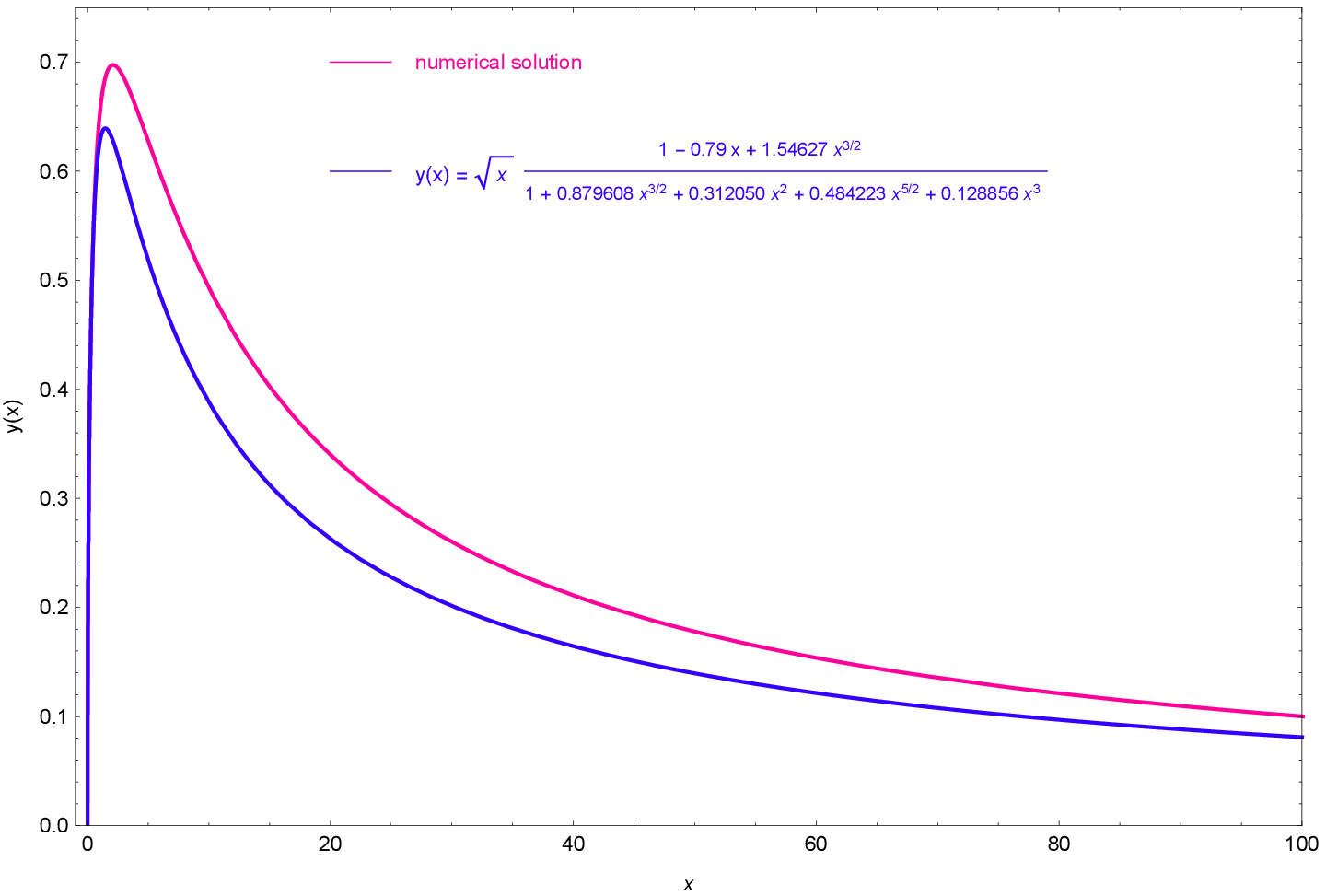}
\caption{Our $y_{1,2}(x)$ in Eq. (4.17) vs. the numerical solution.}
\end{figure}

\begin{figure}
\centering
\includegraphics[height=8cm]{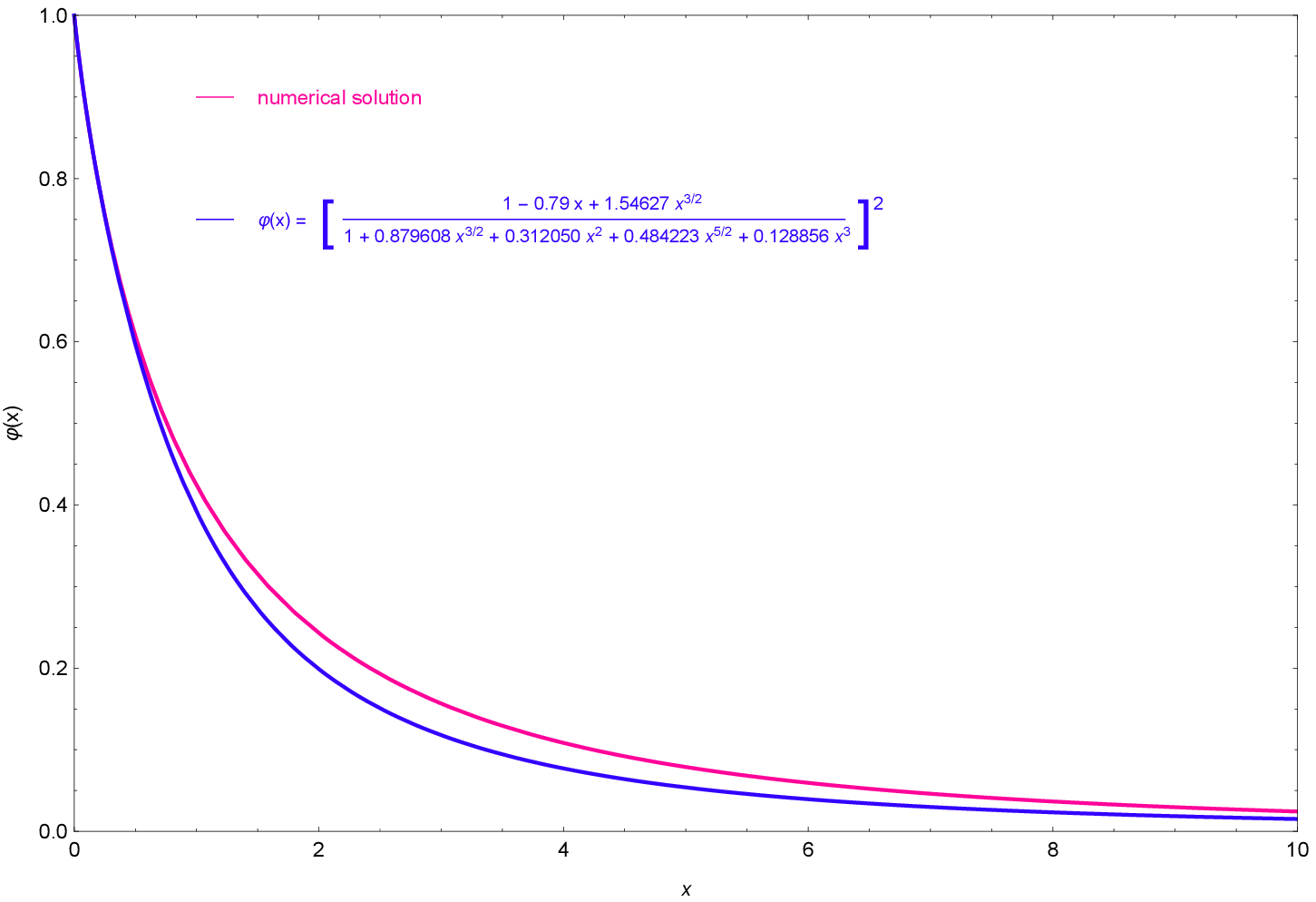}
\caption{Our $\varphi_{1,2}(x)$ vs. the numerical solution.}
\end{figure}

\begin{figure}
\centering
\includegraphics[height=8cm]{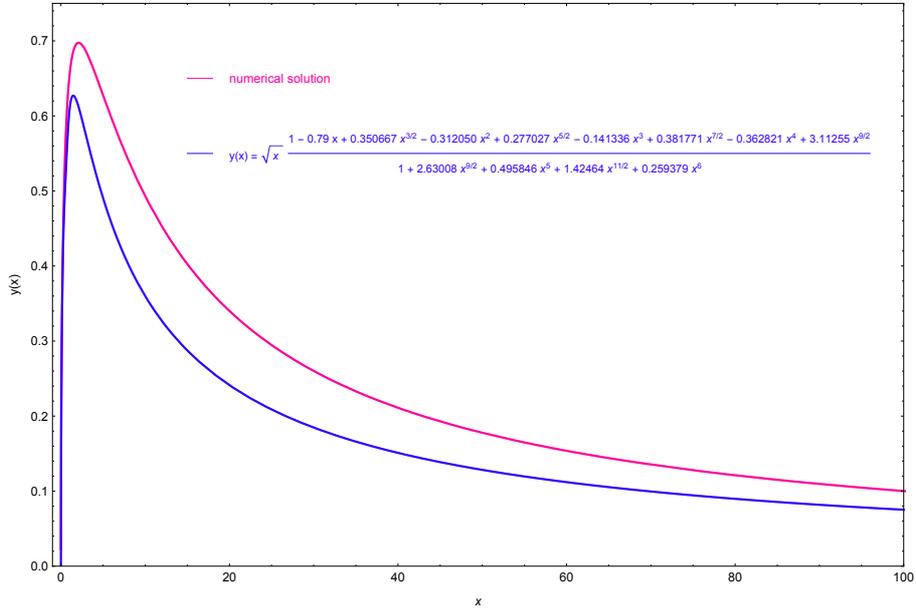}
\caption{Our $y_{5,6}(x)$ from Eq. (4.5) vs. the numerical solution.}
\end{figure}

\begin{figure}
\centering
\includegraphics[height=8cm]{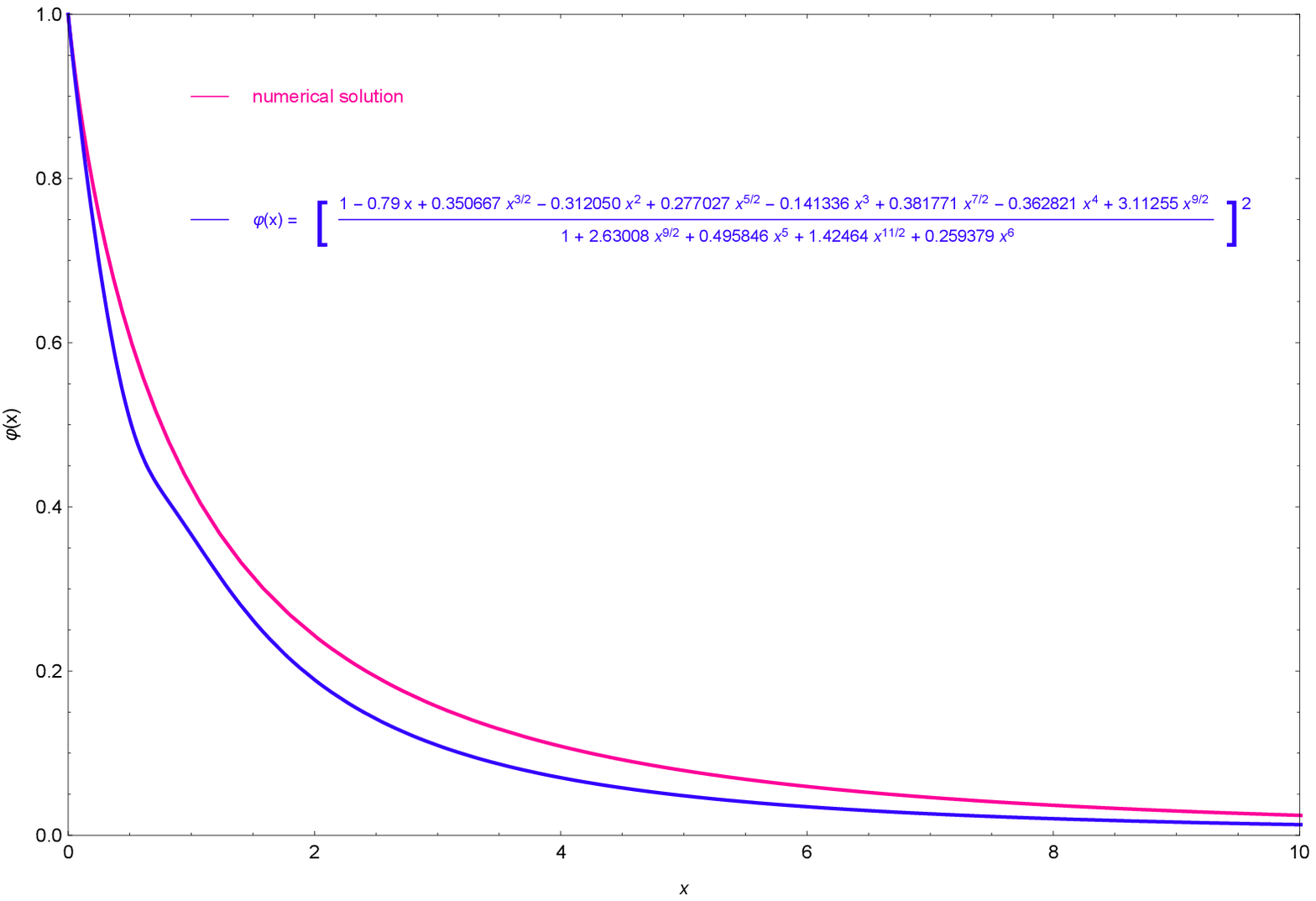}
\caption{Our $\varphi_{5,6}(x)$ vs. the numerical solution.}
\end{figure}

Our findings are as follows.
\vskip 0.3cm
\noindent
(i) Upon adding polynomial terms to numerator and denominator in the
general formulae (4.3) and (4.5), the agreement between our approximate
solutions and the numerical solutions starts worsening. More precisely,
the additional terms lead to deviations from the numerical solutions
at intermediate values of $x$, while for $x \rightarrow 0$ and
$x \rightarrow \infty$ there is still good agreement. However, no
conclusive evidence exists for the need to include or avoid bigger
values of $l$ in Eqs. (4.3) and (4.5).
\vskip 0.3cm
\noindent
(ii) As is clear from our plots, special attention must
be payed to the interval of values of $x$ for which
the denominator in Eqs. (4.16) and (4.17) approaches $0$.
If we focus on $y(x)$ and on its pronounced maximum as the
denominator approaches $0$, we discover that our approximate
solution reproduces well such a feature, but there is no
overlapping with the plot of the numerical solution.

\section{Modern applications}
\setcounter{equation}{0}

Th\-e Tho\-mas-Fer\-mi e\-qua\-ti\-on ha\-s be\-en 
ap\-pli\-ed and ex\-ten\-ded to ma\-ny bran\-ches
of mo\-de\-rn phy\-si\-cs un\-til very recent times, including many-body
systems in quantum mechanics \cite{M1,M2,M3}, semiclassical theory
of atoms \cite{M4}, mathematical refinements \cite{M5,M6,M7},
non-extensive statistical mechanics and relativistic formulation
of the generalized non-extensive Thomas-Fermi model \cite{M7,M8}.
In particular, we are here interested in the differential
equation resulting from the latter framework. The work in Ref.
\cite{M8} has proved that, upon defining the parameters
\begin{equation}
\alpha \equiv \left({4 Z^{2}\over 3 \pi}\right)^{2 \over 3}
{e^{4}\over \hbar^{2}c^{2}},
\label{(6.1)}
\end{equation}
\begin{equation}
a \equiv \left({9 \pi^{2}\over 128 Z}\right)^{1 \over 3}
{\hbar^{2}\over me^{2}},
\label{(6.2)}
\end{equation}
and denoting by $q$ the parameter that measures the departure
of entropy from its additive nature in standard thermodynamics,
one can further define the integrals
\begin{equation}
I_{n}^{(q)} \equiv q \int_{-\infty}^{\infty}
{z^{n} [1+(q-1)z]^{1 \over (q-1)} \over
\left \{ 1+[1+(q-1)z]^{q \over (q-1)} \right \}^{2}} dz,
\label{(6.3)}
\end{equation}
and the parameters ($T$ being the temperature in energy units)
\begin{equation}
\beta \equiv {3 \over 2} T I_{1}^{(q)} 
{a \over e^{2} Z},
\label{(6.4)}
\end{equation}
\begin{equation}
\gamma \equiv {3 \over 8} T^{2} I_{2}^{(q)}
{a^{2}\over e^{4}Z^{2}},
\label{(6.5)}
\end{equation}
so that the desired generalized form of the Thomas-Fermi equation 
(1.2) reads as
\begin{equation}
\sqrt{x}{d^{2}\varphi \over dx^{2}}
=\varphi^{3 \over 2} 
\left[1+\alpha {\varphi \over x}\right]^{3 \over 2}
\left \{1+ \beta {x \over \varphi}
\left[1+\alpha {\varphi \over x}\right]^{-1}
+\gamma {x^{2}\over \varphi^{2}}
\left[1+\alpha {\varphi \over x}\right]^{-2} \right \}.
\label{(6.6)}
\end{equation}
In this equation, relativistic effects are included by means of
the $\alpha$ parameter, while non-extensive and thermal effects
correspond to $\beta$ and $\gamma$, respectively \cite{M8}.
It should be stressed that both non-extensive and thermal corrections
depend on the $q$ parameter, that underpins the generalized entropy
and is linked to the underlying dynamics of the atomic system while
also providing a measure of the degree of its correlation.

At this stage, if we define the function $y$ as in Eq. (2.2),
we obtain eventually the non-linear equation
\begin{equation}
{d^{2}y \over dx^{2}}-{2 \over x}{dy \over dx}
+{y \over x^{2}}
+{1 \over y} \left({dy \over dx}\right)^{2}
={1 \over 2}{y^{2}\over x}
F_{\alpha \beta \gamma}(x,y(x)),
\label{(6.7)}
\end{equation}
having defined
\begin{equation}
f_{\alpha}(x,y(x)) \equiv \left(
1+ \alpha {y^{2}\over x^{2}}\right),
\label{(6.8)}
\end{equation}
\begin{equation}
F_{\alpha \beta \gamma}(x,y(x)) \equiv
f_{\alpha}^{3 \over 2} \left [
1+\beta {x^{2} \over y^{2}} f_{\alpha}^{-1}
+\gamma {x^{4}\over y^{4}}f_{\alpha}^{-2} \right].
\label{(6.9)}
\end{equation}

The desired approximate solution of Eq. (6.7) 
differs substantially from the Sommerfeld solution, as we will show in the following.

\subsection{Small deviations from the Sommerfeld asymptotics}

Let us look for an asymptotic expansion of the solution of the 
Thomas-Fermi equation in the limit $x \rightarrow \infty$. 
First of all, we will explore the inverse power law 
\begin{equation}
y \sim \frac{k}{x^\eta}
\label{(6.10)}
\end{equation}
for large $x$ and positive $\eta$ (corresponding to the physically 
relevant case of a vanishing electrostatic potential at large distances). 
As a first step, let us consider the standard Thomas-Fermi equation 
(\ref{(6.7)}) with $F_{\alpha \beta \gamma} =1$; by substituting 
Eq. (\ref{(6.10)}) into this equation we find:
\begin{equation}
\frac{2(2 \eta^2 + 3 \eta +1)}{x^{\eta+2}} = \frac{k}{x^{2 \eta + 1}} ,
\label{(6.11)}
\end{equation}
which means that Eq. (\ref{(6.10)}) yields a solution of the standard 
Thomas-Fermi equation provided that
\begin{equation}
\left\{ \begin{array}{l}
2(2 \eta^2 + 3 \eta +1) = k ,\\
\eta+2 = 2 \eta + 1 ,
\end{array} \right. \qquad 
\Longrightarrow \qquad 
\left\{ \begin{array}{l}
k = 12 , \\
\eta = 1 .
\end{array} \right.
\label{(6.12)}
\end{equation}
We then find the well-known result that the Sommerfeld solution $y \sim 12/x$ 
is the only inverse power-law solution of the standard Thomas-Fermi equation
at large $x$.

Let us now restore the term $F_{\alpha \beta \gamma} \neq1$ into Eq. 
(\ref{(6.7)}); by substituting Eq. (\ref{(6.10)}), and retaining 
only leading terms for large $x$, so that
\begin{equation}
F_{\alpha \beta \gamma} \sim 1 + \frac{3}{2} \, \alpha \, \frac{k^2}{x^{2(\eta+1)}} 
+ \beta \, \frac{x^{2(\eta+1)}}{k^2} \, + \, \gamma \, 
\frac{x^{4(\eta+1)}}{k^4} \sim \gamma \, \frac{x^{4(\eta+1)}}{k^4} ,
\label{(6.13)}
\end{equation}
we obtain
\begin{equation}
\frac{2 k^3(2 \eta^2 + 3 \eta +1)}{x^{\eta+2}} = \gamma \, {x^{2 \eta + 3}}  .
\label{(6.14)}
\end{equation}
This implies that, in order for Eq. (\ref{(6.7)}) to be satisfied 
(in the large-$x$ limit) we should impose
\begin{equation}
\left\{ \begin{array}{l}
2 k^3 (2 \eta^2 + 3 \eta +1) = \gamma ,\\
- \eta - 2 = 2 \eta + 3 ,
\end{array} \right. 
\label{(6.15)}
\end{equation}
finding therefore a negative $\eta = - 5/3$, which of course does not correspond to 
an inverse power law. This means that non-standard effects in the 
modified Thomas-Fermi equation (\ref{(6.7)}) prevent such a solution for 
large $x$ even, quite interestingly, just approximately for small (but finite) 
$\alpha, \beta, \gamma$ parameters. This result is not unexpected, since it 
results from the divergent part of $F_{\alpha \beta \gamma}$ for $y\sim k/x^\eta$, 
i.e., the last two terms in Eq. (\ref{(6.9)}), that are proportional to $\beta$ 
and $\gamma$. A notable exception is the inclusion of only relativistic 
effects in the Thomas-Fermi equation, for which
\begin{equation}
F_{\alpha 00} \sim 1 + {3 \over 2} \, \alpha \,  {y^{2}\over x^{2}}  .
\label{(6.16)}
\end{equation}
In such a case, however, the Sommerfeld solution $y \sim 12/x$ is only an approximate 
one for small values of the $\alpha$ parameter.

From a strictly physical viewpoint, we expect that the solution of the modified 
Thomas-Fermi equation (with $F_{\alpha \beta \gamma} \neq 1$) tends to that 
of the standard one (with $F_{\alpha \beta \gamma} = 1$) for small values of the 
$\alpha, \beta, \gamma$ parameters. This has to be true also in the large-$x$ 
limit, so that, in such a limit, for $y \sim 12/x$ we should recover 
$F_{\alpha \beta \gamma} \sim 1$ for $\alpha, \beta, \gamma \rightarrow 0$. Now, since 
\begin{equation}
\alpha \,  {y^{2}\over x^{2}} \sim \alpha \,  {144\over x^{4}} \rightarrow 0
\label{(6.17)}
\end{equation}
for $x \rightarrow \infty$ independently of the value of $\alpha$ (so that 
$f_\alpha \rightarrow 1$ for $x \rightarrow \infty$), we should have 
\begin{equation}
\beta \,  {x^{2}\over y^{2}} + \gamma \, {x^{4}\over y^{4}} \rightarrow 0
\label{(6.18)}
\end{equation}
in this limit, or, retaining only the leading term, 
\begin{equation}
\gamma \, x^{4} \rightarrow 0 .
\label{(6.19)}
\end{equation}
Such a term effectively vanishes for $\gamma \rightarrow 0$, {\it provided} that 
the actual value of $x$ is not exceedingly large since, for fixed values of the 
parameter $\gamma$, $x$ might increase indefinitely. In other words, there 
should exist a large but finite value $x_\infty$ for which the asymptotic 
solution $y \sim 12/x$ holds as long as $x \ll x_\infty$. By contrast, notwithstanding 
$\gamma \rightarrow 0$, the condition (\ref{(6.18)}) no longer holds in 
the opposite limit for ever increasing $x$ but, as long as we still approximately 
have $y \sim 12/x$, from the requirement that $F_{\alpha \beta \gamma} \sim 1$ 
for $x \sim x_\infty$ we now find 
\begin{equation}
\gamma \, {x_\infty^{8}\over 12^{4}} \rightarrow 1 .
\label{(6.20)}
\end{equation}
We thus deduce that the cutoff value $x_\infty$,
\begin{equation}
x_\infty \sim \frac{\sqrt{12}}{\sqrt[8]{\gamma}}  ,
\label{(6.21)}
\end{equation}
diverges for $\gamma$ approaching zero, as expected.

Of course, for $x \gg x_\infty$ the asymptotic expression $y \sim 12/x$ 
is no longer valid and non-standard effects strongly affect the 
behaviour of $y(x)$, as we will see below. 

\subsection{Emergence of non-standard effects}

The negative-$\eta$ solution of Eq. (\ref{(6.15)}) would suggest a 
mathematical ansatz $y \sim s \, x^\tau$ with a positive power $\tau$, 
describing only non-standard effects for large $x$. However, following 
the same lines of reasoning as above, it is simple to show that similar 
contradictions as for (\ref{(6.15)}) arise both for $0<\tau<1$ and for 
$\tau >1$, with the interesting exception of the case $\tau =1$. 
Indeed, by substituting
\begin{equation}
y \sim s \, x
\label{(6.22)}
\end{equation}
into equation (\ref{(6.7)}), we find that the modified Thomas-Fermi 
equation is satisfied provided that 
\begin{equation}
0 = \frac{1}{2} \, s^2 x F_{\alpha \beta \gamma}  .
\label{(6.23)}
\end{equation}
Such a condition is actually fulfilled for $x \sim 0$ and any value of the 
non-standard parameters $\alpha, \beta, \gamma$, so that the linear function 
in Eq. (\ref{(6.22)}) is an approximate solution of the modified Thomas-Fermi 
equation (for any value of $s$) in the small-$x$ regime, thus deviating 
appreciably from the $\sqrt{x}$ behaviour of the standard Thomas-Fermi 
case studied earlier.

By contrast, for finite $x$ (and $s$), from definition 
(\ref{(6.9)}) the condition (\ref{(6.23)}) 
leads to the requirement that the constant $s$ should satisfy the relation
\begin{equation}
\left( 1 + \alpha \, s^2 \right)^{3/2} \left[ 1 + \frac{\beta}{s^2 (1 
+ \alpha \, s^2)} + \frac{\gamma}{s^4 (1 + \alpha \, s^2)^2} \right] = 0  ,
\label{(6.24)}
\end{equation}
which displays a physically realizable solution. Indeed, 
for any value of the non-standard (positive) parameters 
$\alpha, \beta, \gamma$, we obtain $s=s_\ast$, with
\begin{equation}
s_\ast = \sqrt{\frac{1}{2 \alpha} \left[ \sqrt{1+ 2 \alpha \left( 
\sqrt{\beta^4 + 4 \gamma^2} - \beta \right)} - 1 \right]} .
\label{(6.25)}
\end{equation}
Thus, the linear solution (\ref{(6.22)}) effectively rules strong 
non-standard effects, for which $F_{\alpha \beta \gamma}$ 
vanishes rather than approaching the unit value.

\subsection{Small-$x$ solutions}

Non-standard effects resulting from a linear behaviour (underlying a vanishing 
$F_{\alpha \beta \gamma}$ term) for finite $x$ are even more pronounced in the 
neighborhood of the origin. This can be explored by looking for 
an approximate solution in the form
\begin{equation}
y \sim Q \, x^{\tau} ,
\label{(6.26)}
\end{equation}
for $x \rightarrow 0$ and positive $\tau$. It is straightforward to see that 
in the small $x$-regime, for $\tau <1$, we recover the square root behaviour 
(corresponding to $\tau = 1/2$) for vanishing values of the $\alpha$ parameter, 
by simply substituting Eq. (\ref{(6.26)}) into the modified Thomas-Fermi 
equation (\ref{(6.7)}). However, for finite values of the non-standard parameters 
(mainly ruled by $\gamma$, as we will see), the small-$x$ behaviour 
manifests itself into a different, larger value of the exponent. 
Again, substitution of Eq. (\ref{(6.26)}) into Eq. (\ref{(6.7)}) 
leads for $\tau >1$ to the equation
\begin{equation}
\frac{2 Q^3 (2 \tau^2 - 3 \tau + 1)}{x^{2-\tau}} = \frac{\gamma}{x^{2 \tau -3}}  ,
\label{(6.27)}
\end{equation}
when retaining only the leading terms, which is actually satisfied provided that
\begin{equation}
\left\{ \begin{array}{l}
\displaystyle 2 Q^3 (2 \tau^2 - 3 \tau +1) = \gamma ,\\
\displaystyle 2 - \tau = 2 \tau - 3 ,
\end{array} \right. \qquad 
\Longrightarrow \qquad 
\left\{ \begin{array}{l}
\displaystyle Q = Q_\ast \equiv \sqrt[3]{\frac{9 \gamma}{28}} \, , \\ \\
\displaystyle \tau = \frac{5}{3}  .
\end{array} \right.
\label{(6.28)}
\end{equation}
Note that, in the small-$x$ regime, such a solution corresponds to a 
large value of the $F_{\alpha \beta \gamma}$ term in the modified Thomas-Fermi 
equation that, although signaling a non-standard behaviour ($F_{\alpha \beta \gamma} \neq 1$), 
is at variance with the linear solution underpinning a vanishing 
$F_{\alpha \beta \gamma}$ term. However, as for the linear case, such 
$x^{5/3}$ behaviour is again ruled by thermal effects by means of a finite 
value of the $\gamma$ parameter.

\section{Concluding remarks}
\setcounter{equation}{0}

Our sections 2-5 have been devoted to a detailed investigation of the
non-relativistic Thomas-Fermi boundary-value problem. As far as we
know, our auxiliary differential equation (2.6), the form
(3.4)-(3.6) of the boundary conditions, and the two families of 
approximate solutions in Eq. (4.10) are completely new, as well
as the particular examples in Eqs. (4.16) and (4.17). The fairly 
good agreement with the numerical solution, displayed in the plots
of Section 5, is encouraging. Moreover, the smooth transition from the
small-$x$ to the large-$x$ behaviour is another merit of our original
approximate solutions.

In section $6$, we have instead investigated the joint effect of
relativistic, non-extensive and thermal effects in the Thomas-Fermi
equation, and we have discovered the following approximate power-law behaviours 
for the solution of the modified Thomas-Fermi equation in the different regimes:
\[
\begin{array}{rlll}
{\rm small} \, x: & \quad y \sim k \, \sqrt{x}, & \quad {\rm small} \, 
\alpha,  & \quad {\rm arbitrary} \, F_{\alpha \beta \gamma}; 
\\ \\ 
{\rm small} \, x: & \quad \displaystyle y \sim Q_\ast \, x^{5/3}, & \quad 
{\rm finite} \, \alpha, \beta, \gamma,  & \quad {\rm large} \, F_{\alpha \beta \gamma}; 
\\ \\ 
{\rm finite} \, x: & \quad \displaystyle y \sim s_\ast \, x, & \quad {\rm finite} \, 
\alpha, \beta, \gamma,  & \quad {\rm small} \, F_{\alpha \beta \gamma}; 
\\ \\ 
\begin{array}{r} {\rm large} \, x : \\ \left[ x \ll x_\infty \right] \end{array} 
\!\! & \quad \displaystyle y \sim \frac{12}{x}, & \quad {\rm small} \,\beta, \gamma,  
& \quad {\rm unit} \, F_{\alpha \beta \gamma}; 
\end{array}
\]
The original calculations of our work provide encouraging evidence in 
favour of the Thomas-Fermi equation being a valuable source of inspiration
for understanding the wide range of
modern applications \cite{M9,M10,M11,M12,M13} of atomic physics.

\section*{Acknowledgments}
The authors are grateful to the Dipartimento di Fisica 
``Ettore Pancini'' of Federico II University for hospitality 
and support.

\end{document}